 \documentclass[final,5p,times,twocolumn]{elsarticle}

\usepackage{graphics}
\usepackage{epstopdf}
\usepackage{amsmath}
\usepackage{amssymb}

\journal{Physics of the Dark Universe}

\begin{document}

\begin{frontmatter}

\title{CFT Duals on Rotating Charged Black Holes Surrounded by Quintessence}

\author[label1,label2]{Muhammad F. A. R. Sakti\corref{cor1}}
\address[label1]{Theoretical Physics Laboratory, THEPi Division, Institut Teknologi Bandung, Jl. Ganesha 10 Bandung, 40132, Indonesia}
\address[label2]{Indonesia Center for Theoretical and Mathematical Physics (ICTMP), Institut Teknologi Bandung, Jl. Ganesha 10 Bandung, 40132, Indonesia}

\cortext[cor1]{Corresponding author}

\ead{fitrahalfian@gmail.com}

\author[label1,label2]{Freddy P. Zen}
\ead{fpzen@fi.itb.ac.id}

\begin{abstract}
We demonstrate that Kerr/CFT duality can be extended to rotating charged black hole solution surrounded by quintessence in Rastall gravity. Since Rastall gravity can be considered as $F(R,T) =  R + \beta T $ theory with the addition of a matter term, so the gravitational Lagrangian is resemblant to Einstein general relativity. In fact, the resulting central term is coincident to the central term in the general relativity since $ T $ is only a trace of the matter term. We then exhibit that the entropy from CFT of this black hole is equivalent with the Bekenstein-Hawking entropy with no correction related to the Rastall coupling constant. Moreover, when Rastall coupling constant $ \kappa \lambda $ is switched off and $ a $ is approaching zero, we may extend the solution to the 5D black hole solution. It is found that this extremal solution contains twofold CFT duals that we call as $Q$-picture and $ P $-picture. In both pictures, the CFT entropy reproduces exactly the Bekenstein-Hawking entropy for Reissner-Nordstr{\"o}m black hole surrounded by quintessence. The results establish that, for all black holes surrounded by quintessence, the microscopic entropies of the dual CFTs agree with the Bekenstein-Hawking entropies.\end{abstract}

\begin{keyword}
Rastall gravity, quintessence, thermodynamics, Kerr/CFT correspondence
\end{keyword}

\end{frontmatter}


\section{Introduction}
In the late 1990s, astronomers have discovered the evidence that the expansion of the universe is not slowing down due to the gravity as expected. Instead, the expansion of the universe is accelerated. This acceleration is believed to be powered by an unknown nature called as dark energy \cite{Copeland2006}. High-precision observational data show that dark energy makes up around 70\% of the universe. Dark energy commonly is modeled as a matter of a perfect fluid equation of state (EoS) parameterized by a constant $ \omega $. The value of $ \omega $ can be in the range $ -1 <\omega <  -1/3 $ for an accelerated expansion of the universe and called as quintessence. Quintessence is one of the simplest models for it. Yet, that is also studied as an ingredient for some compact objects \cite{Kiselev2003,Maharaj2017,Lee2018,SaktiarXivString2019}.

Einstein general relativity assumes that the conservation law of the energy-momentum tensor is satisfied, $ \nabla_\mu T^{\mu\nu} =0 $. However, this basic circumstance is deemed to be violated for Rastall gravitational theory. Rastall has assumed that there can be a deviation in the energy-momentum tensor conservation parameterized by a Rastall coupling constant $ \lambda $ (or sometimes $ \kappa\lambda $), i.e.$ \nabla_\mu T^{\mu\nu} \sim \lambda a_\mu $ where $ a_\mu $ is a rank-1 tensor which is deemed to be $ \nabla_\mu R $ \cite{Rastall1972,Rastall1976}. In flat space-time, we know that $ R=0 $, so Rastall theory is still feasible. In addition, in a gravitational system, quantum effects may lead to the violation of the conservation of the energy-momentum. In fact, Rastall gravity can be considered as a phenomenological way of determining the effects of quantum fields in the curved space–time in a covariant way. Unlike the original formulation, it is argued in \cite{Fisher2019,Moraes2019,ShabaniZiaieEPL2020} that Rastall gravity can be derived from a specific condition of $ F(R,T) $ theory with the addition of a matter term where this leads to the non-conservative covariant derivative of the energy-momentum tensor (see \ref{app:Rastalltheor} for the derivation). Rastall theory has also been applied in the cosmological scale \cite{Campos2013,Fabris2012,Moradpour2016,Rawaf1996,Batista2012,Carames2014,Salako2016,Smalley1974,Smalley1975,Wolf1986,Moradpour2017} to some astrophysical compact objects \cite{Hansraj2019,Abbas2018,Abbas2019,Oliveira2015,Moradpour2017a,Moradpour2016a,Oliveira2016,Bronnikov2016,SaktiAnnPhys2020,HadyanIJMPD2020}. 

It is conjectured that from the quantum states in the near-horizon region of the Kerr black hole, we can identify a certain two-dimensional (2D) chiral conformal field theory (CFT) \cite{Guica2009}. The identification of the 2D chiral CFT has been carried out by examining the asymptotic symmetry group with some arbitrary boundary conditions on the boundary where CFT lives that will generate a class of diffeomorphisms of the near-horizon Kerr geometry.  By defining charges associated with the transformations and evaluating the Dirac brackets of the charges, a Virasoro algebra is produced with the presence of a quantum anomaly proportional to the angular momentum of the black hole. In addition, the calculation of the temperature is also needed by assuming Frolov-Thorne vacuum, the generalized Hartle-Hawking vacuum. The last and important stage is the calculation of the microscopic entropy of Kerr black hole using the Cardy's growth of states for a 2D CFT. The calculation indicates that the resulting microscopic entropy matches precisely with the Bekenstein-hawking entropy. These results establish that the extremal Kerr black hole in the near-horizon region is holographically dual with 2D CFT. This computation has also been extended to many black hole solutions \cite{Hartman2009, Ghezelbash2009, Lu2009, Li2010, Anninos2010, Ghodsi2010, Ghezelbash2012, Astorino2015, Astorino2015a, Siahaan2016, Astorino2016, Sinamuli2016, Sakti2018, SaktiEPJPlus2019, SaktiMicroJPCS2019} and also to the generic rotating black holes where the space-time metric is used as a background of the quantum field \cite{Castro2010, ChenLong2010, ChenSun2010, ChenWang2010, WangLiu2010, ChenLongJHEP2010, SetareKamali2010, GhezelbashKamali2010, ChenHuangPRD2010, ChenChen2011, ChenGhezelbash2011, ChenZhangJHEP2011, HuangYuan2011, Shao2011, DeyouChen2011, GhezelbashSiahaan2013, SiahaanAcc2018, Saktideformed2019, SaktiNucPhysB2020}. For the review, one can see \cite{Compere2017}.

Numerous black hole solutions have been derived in Rastall gravity since this gravitational theory is one of promising general gravitational theories. An example of black hole solutions in Rastall gravity is a rotating charged and twisting black hole surrounded by quintessence in Rastall gravity \cite{SaktiAnnPhys2020}. In this paper, we wish to perform the conjectured Kerr/CFT duality to the rotating charged black hole surrounded by quintessence in Rastall gravity derived in \cite{SaktiAnnPhys2020} but with vanishing NUT charge. Since Rastall gravity can be constructed from $ F(R,T) = R + \beta T $ theory with the addition of a matter term \cite{ShabaniZiaieEPL2020}, we can infer that the information of the central charge is stored only in the space-time metric. Hence, the contribution of all matters is accommodated in the space-time metric. We perform that this black hole is holographically dual to CFT represented by left-moving central charge. Moreover, as the 5D Reissner-Nordstr{\"o}m solution \cite{Ghodsi2010}, we extend the calculation to compute the entropy of 5D Reissner-Nordstr{\"o}m solution surrounded by quintessence. This extension has been performed in several papers when $ a\rightarrow 0 $ \cite{Hartman2009,Sakti2018,SaktiEPJPlus2019}. It is fascinating that we can construct 5D solutions in two different pictures with the fact that there is dual electromagnetic potential. Our computation exhibits that there is twofold CFT dual that reproduces precisely the Bekenstein-Hawking entropy of Reissner-Nordstr{\"o}m solution surrounded by quintessence from Cardy entropy formula.

We set up the remaining parts of the paper as the following. The second section is to briefly review the rotating charged black hole surrounded by quintessence in Rastall gravity. In the section 3, we perform the entropy calculation using the Kerr/CFT duality. Furthermore, we show the entropies for several specific values of $ \kappa\lambda, \omega $ in the section 4. Next, the twofold CFT dual of extremal Reissner-Nordstr{\"o}m surrounded by quintessence is examined. In the concluding section, we give our conclusions of the whole calculation that we perform.

\section{Black Hole Surrounded by Quintessence in Rastall Gravity}\label{KiselevRastallBH}

In this section, we wish to briefly review the rotating charged black hole surrounded by quintessence in Rastall gravity. The space-time metric of this solution is given by \cite{SaktiAnnPhys2020},
\begin{eqnarray}
ds^2 &=& - \frac{\Delta}{\varrho^2} \bigg( dt - a \sin^2\theta d\phi \bigg)^2 +  \frac{\varrho ^2}{\Delta}d\hat{r}^2 + \varrho ^2 d\theta^2 \nonumber\\
& &+ \frac{\sin^2\theta}{\varrho ^2}\bigg[adt - (\hat{r}^2+a^2 ) d\phi \bigg]^2,\label{KNQRmetric} \
\end{eqnarray}
where
\begin{equation}
\Delta =\hat{r}^2- 2M \hat{r} +a^2 + q^2 +p^2-\alpha \hat{r}^\upsilon ,\label{Del} \nonumber\
\end{equation}
\begin{equation}
\upsilon = \frac{1-3\omega}{1-3\kappa\lambda(1+\omega)}, ~~~ \varrho^2 =\hat{r}^2+ a^2\cos^2\theta .\label{rho} \nonumber\
\end{equation}
This solution contains a Rastall coupling constant $ \kappa\lambda $. The parameters $a$ and $M$ represent the rotation parameter and mass, respectively. The quintessential intensity, equation of state (EoS) parameter, electric charge and magnetic charge of the space-time are given by $ \alpha $, $\omega$, $q$ and $p$, respectively. The space-time metric (\ref{KNQRmetric}), in the limit of $a=0$, reduces to the Reissner-Nordstr{\"o}m black hole surrounded by quintessence in Rastall gravity. Moreover, it reduces to Schwarzschild space-time surrounded by quintessence in the limit of $a,q,p,\kappa\lambda=0$. It is interesting to add the cosmological constant in the metric (\ref{KNQRmetric}) like the authors in \cite{XuBHRastallAdS2018} have done. However, we find that their solution does not reduce to the Reissner-Nordstr{\"o}m-AdS black hole surrounded by quintessence in Rastall gravity (see \ref{app:KNUTAdSQ}). 

The electromagnetic potential related to the metric (\ref{KNQRmetric}) is given by  \cite{SaktiAnnPhys2020}
\begin{equation}
\textbf{A} = \frac{-q\hat{r}\left[a dt- a^2 \sin^2\theta  d\phi \right]}{a\varrho ^2} - \frac{p \text{cos}\theta \left[a dt- \left( \hat{r}^2 + a^2 \right) d\phi \right]}{\varrho ^2} .\ \label{eq:electromagneticpotentialKNQR}
\end{equation}
The dual electromagnetic potential is then given by
\begin{equation}
\bar{\textbf{A}} = \frac{-p\hat{r}\left[a dt- a^2 \sin^2\theta  d\phi \right]}{a\varrho ^2} + \frac{q \text{cos}\theta \left[a dt- \left( \hat{r}^2 + a^2 \right) d\phi \right]}{\varrho ^2} .\ \label{eq:dualelectromagneticpotentialKNQR}
\end{equation}
The dual electromagnetic potential satisfies
\begin{equation}
\bar{\textbf{F}} =d\bar{\textbf{A}}, ~~\text{such~that}~~ \bar{\textbf{F}} = *\textbf{F}.
\end{equation}
Sign $ * $ is the Hodge dual.

Some thermodynamic properties of this solution have been derived in \cite{SaktiAnnPhys2020}. The electric potential and the general force related to the quintessence are given as follows
\begin{eqnarray}
&&\Phi_q =\frac{qr_+}{r_+^2+a^2},\label{en4KNQR}\\
&&\Theta_H =-\frac{1}{2}r_+^{\upsilon -1}.\label{en6KNQR}\
\end{eqnarray} 
The general force of the quintessence has been proposed originally in \cite{ChenPRDQuin2008} as we can also consider the cosmological constant as a thermodynamic quantity \cite{Sekiwa2006}. From the electromagnetic duality, the magnetic potential is given by
\begin{equation}
\Phi_p =\frac{pr_+}{r_+^2+a^2} .\label{en5KNQR}\
\end{equation}
The Bekenstein-Hawking entropy, Hawking temperature, and angular velocity on the event horizon $r_+$ of the black hole,
are given by
\begin{eqnarray} \label{enKNQR}
&&S_{BH}=\frac{A_{BH}}{4}=\pi\left(r_+^2+a^2\right),\label{en1}\\
&&T_H=\frac{2(r_+ -M) -\alpha\upsilon r_+^{\upsilon -1}}{4\pi(r_+^2+a^2)},\label{en2KNQR}\\
&&\Omega_H =\frac{a}{r_+^2+a^2},\label{en3KNQR}\
\end{eqnarray} 
respectively. As we can see, the entropy derived in \cite{SaktiAnnPhys2020} as we show above is equal to the quarter of the black hole's area. This result is coincident with the macroscopic computation of the entropy given in \cite{ChakrabortyRastallEnt2018} that proposes that the horizon entropy in the Rastall gravity is similar with Einstein general relativity. Nevertheless, this result is in contrast with the result of \cite{MoradpourRastallEnt2016} where there is a correction that is dependent on Rastall parameter where the total entropy is given as follows
\begin{equation}
S =\frac{A_{BH}}{4}+\frac{A_{BH}}{4}\left(\frac{2\kappa\lambda}{4\kappa\lambda -1} \right), \label{eq:Scorrected}
\end{equation}
where the second term on the right hand side will vanish in Einstein gravity. In order to obtain the Bekenstein-Hawking entropy as a horizon entropy in Rastall gravity, the authors in \cite{ChakrabortyRastallEnt2018} have modified the Clausius relation that leads to the Misner-Sharp mass of the Einstein theory to obtain the Bekenstein-Hawking entropy. It is worthwhile pointing out that we need to modify the Clausius relation related to the theories, for instance in Lovelock gravity and Brane-World scenario \cite{ChaiZhangWangCTP2005,ChakraPadmaPRD2014a,ChakraPadmaPRD2014b,MitraGRG2015}. 

Regarding the entropy calculation, we will derive the entropy of the black holes surrounded by quintessence both in Einstein and Rastall gravities by implementing the Kerr/CFT correspondence. In this manner, the famous Cardy formula will be implemented to compute the entropy. For the solution in Rastall gravity, we can identify whether the resulting entropy is similar with the Bekenstein-Hawking entropy (\ref{enKNQR}) or the Bekenstein-Hawking entropy with the correction (\ref{eq:Scorrected}) to prove the result given in \cite{ChakrabortyRastallEnt2018}.

\section{Entropy Calculation from CFT}\label{entropy calculation}
In this section, we wish to derive the entropy for the black hole solution surrounded by quintessence using the conjectured Kerr/CFT correspondence. Regarding that, we need to examine the excitation around near-horizon extremal black hole solution that will be derived here.

\subsection{Near-horizon Geometry}

To find the near-horizon geometry of the extremal Kerr-Newman black hole surrounded by quintessence (\ref{KNQRmetric}),  
we consider the following coordinate transformations \cite{Hartman2009,Compere2017}
\begin{eqnarray}
\hat{r} = r_+ +\epsilon r_0 r,~~~ t = \frac{r_0}{\epsilon}\tau, ~~~ \phi = \varphi + \frac{\Omega_H r_0}{\epsilon}\tau, \label{extremaltransformation} \
\end{eqnarray}
where $r_0=\sqrt{{r_+^2+a^2}}$. $ \epsilon \rightarrow 0 $ denotes the near-horizon limit. In the near-horizon of extremal black holes, the function $ \Delta $ takes form \cite{SaktiEPJPlus2019}
\begin{equation}
\Delta = (\hat{r}-r_+)^2 V + \mathcal{O}\left((\hat{r}-r_+)^3 \right),\
\end{equation}
where the constant $V$ is given by
\begin{equation}
V = 1 - \frac{\alpha}{2}\upsilon (\upsilon -1)r_+^{\upsilon -2},\
\end{equation}
for the quintessential black hole solution.  Now, metric (\ref{KNQRmetric}) transforms to
\begin{eqnarray}
ds^2 &=& \Gamma(\theta)\left(-V^2 r^2 d\tau^2 + \frac{dr^2}{r^2} + \alpha(\theta) d\theta ^2 \right) \nonumber\\
& & +\gamma(\theta) \left(d\varphi +\frac{2ar_+}{r_0^2}r d\tau\right)^2, \label{extremalmetric}\
\end{eqnarray}
where
\begin{equation}
\Gamma(\theta)= \frac{\varrho_+ ^2}{V} , ~~~\alpha(\theta) = V, ~~~\gamma(\theta)=\frac{r_0^4 \sin^2\theta}{\varrho_+^2}, ~~~
\varrho_+^2 = r_+^2 + a^2\cos^2\theta. \nonumber\
\end{equation}
Using the scaling $ d\tau \rightarrow  d\tau/V $, we obtain
\begin{eqnarray}
ds^2 &=& \Gamma(\theta)\left(-r^2 d\tau^2 + \frac{dr^2}{r^2} + \alpha(\theta) d\theta ^2 \right) \nonumber\\
& &+\gamma(\theta) \left(d\varphi +e r d\tau\right)^2, \label{eq:extremalmetric}\
\end{eqnarray}
where\vspace{-0.5cm}
\begin{equation}
e = \frac{2ar_+}{r_0^2 V}. \label{eq:p}
\end{equation}
In the near-horizon limit, we find that the gauge field (\ref{eq:electromagneticpotentialKNQR}) and its dual are given by
\begin{equation}
\textbf{A} = \frac{r_0^2\left[q(r_+^2-a^2\cos^2\theta \right) +2par_+ \cos \theta]}{2ar_+ \rho_+ ^2} \left(d\varphi +e rd\tau \right) ,\label{eq:nearhorizonelectromagneticpot1}\end{equation}
\begin{equation}
\bar{\textbf{A}} = \frac{r_0^2\left[p(r_+^2-a^2\cos^2\theta \right) -2qar_+ \cos \theta]}{2ar_+ \rho_+ ^2} \left(d\varphi +e rd\tau \right) . \label{eq:nearhorizonelectromagneticpot2}
\end{equation}
The near-horizon space-time metric (\ref{eq:extremalmetric}) resembles the near-horizon metric for more general black hole solutions explained in \cite{Compere2017}. It has an isometry that is generated by the following Killing vectors
\begin{equation}
\zeta_0 = \partial_\varphi , \nonumber\
\end{equation}
\begin{equation}
X_1 = \partial_\tau, ~~~ X_2 = \tau \partial_\tau - r \partial_r, ~~~ X_3 = \left(\frac{1}{2r^2}+\frac{\tau^2}{2} \right)\partial_\tau -\tau r \partial_r - \frac{e}{r}\partial_\varphi .\label{eq:isometrynearhorizon}
\end{equation}
The Killing vectors (\ref{eq:isometrynearhorizon}) denote an enhanced $ SL(2,R) \times U(1) $ isometry group. This isometry is similar with the isometry of $ AdS_2 $ space-time and also a hint to employ the asymptotic symmetry group (ASG) as it has been used by Brown and Henneaux \cite{brown1986} to calculate the central charge of the $ AdS_3 $ space-time.

\subsection{Central Charge}
We perform the approach of Brown and Henneaux \cite{brown1986} to find the central charge of the holographic dual CFT description of this rotating black hole. To calculate the charge that associates with ASG of near-horizon extremal solution (\ref{eq:extremalmetric}), we only consider the contribution of gravitational field and use the formalism in \cite{BarnichBrandt2002}. As noted in \cite{Hartman2009,SaktiEPJPlus2019,Compere2017,CompereMurataJHEP2009}, contributions from matter's Lagrangian will vanish. As shown in \ref{app:Rastalltheor}, Rastall gravity can be constructed from $ \mathcal{L} = R + \beta T + \mathcal{L}_m $ where $ T $ is just the trace of matter stress tensor, so the information of the central charge is stored in the space-time metric only. The asymptotic symmetry of this solution includes the diffeomorphism $ \xi $ such that
\begin{eqnarray}
\delta_\xi g_{\mu\nu} = \mathcal{L}_\xi g_{\mu\nu}.\ 
\end{eqnarray}
The associated conserved charge is given as follows
\begin{eqnarray}
Q_{\xi} = \frac{1}{8\pi}\int_{\partial\Sigma} k^{g}_{\zeta}[h;g],
\end{eqnarray}
where above integral is over the boundary of a spatial slice. The expression for the contribution of the metric tensor on the central charge is then
\begin{eqnarray}
k^{g}_{\zeta}[h;g] &=& -\frac{1}{4}\epsilon_{\rho\sigma\mu\nu} \left\{ \zeta ^{\nu} D^{\mu} h - \zeta ^{\nu} D_{\lambda} h^{\mu \lambda} + \frac{h}{2} D^{\nu}\zeta^{\mu} \right. \nonumber\\
& & - h^{\nu \lambda} D_{\lambda}\zeta^{\mu} + \zeta_{\lambda}D^{\nu}h^{\mu \lambda} +\frac{h^{\lambda \nu}}{2} \left(D^\mu \zeta_\lambda \right. \nonumber\\
& & \left. + D_\lambda \zeta^\mu \right) \bigg\} dx^\rho \wedge dx^\sigma . \label{eq:kgrav} \
\end{eqnarray}
The charge $ Q_{\zeta} $ generates symmetry through the Dirac brackets. Algebra of ASG is given by the Dirac bracket algebra of these charges
\begin{equation}
\{ Q_{\zeta},Q_{\bar{\zeta}} \}_{DB} = \frac{1}{8\pi}\int k^{g}_{\zeta}\left[\mathcal{L}_{\bar{\zeta}}g;g \right]  = Q_{[\zeta,\bar{\zeta}]} + \frac{1}{8\pi}\int k^{g}_{\zeta}\left[\mathcal{L}_{\bar{\zeta}}\bar{g};\bar{g}\right]. \label{eq:charges}\
\end{equation}

We are to specify the boundary conditions on the metric deviations to produce a finite charge. Hence, we adopt the following boundary conditions
\begin{eqnarray}
h_{\mu \nu} \sim \left(\begin{array}{cccc}
\mathcal{O}(r^2) & \mathcal{O}\left(\frac{1}{r^2}\right) &  \mathcal{O}\left(\frac{1}{r}\right) &  \mathcal{O}(1) \\
 &  \mathcal{O}\left(\frac{1}{r^3}\right) &  \mathcal{O}\left(\frac{1}{r^2}\right) &  \mathcal{O}\left(\frac{1}{r}\right) \\
 &  & \mathcal{O}\left(\frac{1}{r}\right) &  \mathcal{O}\left(\frac{1}{r}\right)\\
 &  &  &  \mathcal{O}(1)\
\end{array} \right), \label{eq:gdeviation}
\end{eqnarray}
in the basis $ (\tau,r,\theta, \varphi) $. We then require an additional boundary condition to focus on the central charge that comes from the left-moving Virasoro algebra which is $ Q_{\partial_{\tau}} = 0 $. By employing all boundary conditions (\ref{eq:gdeviation}), the most general diffeomorphism symmetry is produced which is generated by the following vector field
\begin{eqnarray}
\zeta &=& \left\{c_{\tau} + \mathcal{O}\left(r^{-3}\right) \right\}\partial _{\tau} + \left\{-r\epsilon '(\varphi) + \mathcal{O}(1) \right\}\partial _r \nonumber\\
& & + \mathcal{O}\left(r^{-1}\right)\partial _\theta  + \left\{\epsilon(\varphi) + \mathcal{O}\left(r^{-2}\right) \right\}\partial _\varphi ,\
\end{eqnarray}
where $ c_{\tau} $ is an arbitrary constant and the prime $ (') $ denotes the derivative with respect to coordinate $ \varphi $. The vector field includes one copy of the conformal group of a circle which is generated by
\begin{eqnarray}
\zeta_\epsilon = \epsilon(\varphi)\partial_\varphi -r\epsilon '(\varphi)\partial_r .\label{eq:killingASG}
\end{eqnarray}
Under the rotation $ \varphi \sim \varphi+2\pi $, the coordinate $ \varphi $ is periodic. Consequently, we can define $ \epsilon_n = -e^{-in \varphi} $ and $ \zeta_\epsilon =\zeta_\epsilon(\epsilon_n ) $. This identification yields
\begin{eqnarray}
\zeta_\epsilon = -e^{-in \varphi}\partial _\varphi - inre^{-in \varphi}\partial_r . \label{eq:killingASG1}
\end{eqnarray}
By the Lie bracket, the symmetry generator (\ref{eq:killingASG1}) satisfies the following Virasoro algebra
\begin{eqnarray}
i[\zeta_m, \zeta_n]_{LB} = (m-n)\zeta_{m+n},
\end{eqnarray}
without the central term. By defining
\begin{eqnarray}
Q_{\zeta} \equiv L_{n} - \chi \delta_{n,0},
\end{eqnarray}
on Eq. (\ref{eq:charges}), we obtain the conserved charges algebra in quantum version, such that
\begin{equation}
\left[L_m, L_n \right] = (m-n) L_{m+n} + \frac{ar_+}{V} m (m^2-1)\delta_{m+n, 0}, 
\end{equation}
where $ \chi $ is just a constant. We can read off the value of the left-moving central charge as
\begin{eqnarray}
c_L = \frac{12ar_+}{V}. \label{eq:cL}\
\end{eqnarray}

\subsection{CFT Temperatures}
In the following, we compute the temperatures from  CFT in order to apply Cardy entropy formula from 2D CFT. We can employ the Frolov-Thorne vacuum, the generalization of the Hartle-Hawking vacuum. We start with the first law of black hole thermodynamics
\begin{equation}
T_H dS = dM - \Omega_H dJ - \Phi_q dQ_q - \Phi_p dQ_p - \Theta_H Q_\alpha ,
\end{equation}
and the extremal condition that satisfies
\begin{equation}
T_H^{ex} dS = dM - \Omega_H^{ex} dJ - \Phi_q^{ex} dQ_q - \Phi_p^{ex} dQ_p - \Theta_H Q^{ex}_\alpha=0 .
\end{equation}
Therefore we can write
\begin{eqnarray}
T_H dS &=& -\left[ (\Omega_H - \Omega_H^{ex})dJ + (\Phi_e - \Phi_q^{ex}) dQ_q \right.  \nonumber\\
& &\left. + (\Phi_p - \Phi_p^{ex})dQ_g - (\Theta_H - \Theta_h^{ex})dQ_\alpha \right]. \label{eq:constrainofthermo}
\end{eqnarray}
For such constrained variation (\ref{eq:constrainofthermo}), we may write
\begin{eqnarray}
dS = \frac{dJ}{T_L}+\frac{dQ_q}{T_q}+\frac{dQ_p}{T_p} + \frac{dQ_\alpha}{T_\alpha} .\
\end{eqnarray}

Furthermore, a quantum scalar field with eigenmodes of the asymptotic energy $ E $ and angular momentum $ J $ is considered, which is given by as follows
\begin{eqnarray}
\tilde{\Phi} = \sum_{E,J,s} \tilde{\phi} _{E,J,s} e^{-i E t + i J \phi} f_s(r,\theta),
\end{eqnarray}
for a Kerr black hole. However, we need this in the near-horizon form which is then given by
\begin{eqnarray}
e^{-i E t + i J \phi} = e^{-in_R \tau + in_L \varphi},
\end{eqnarray}
where
\begin{eqnarray}
n_R = (E-\Omega_H^{ex} J)r_0/\lambda, ~~~n_L = J. \label{eq:nrnl}
\end{eqnarray}
For a rotating charged black hole surrounded by quintessence, we may extend Eq. (\ref{eq:nrnl}) to 
\begin{equation}
n_R = (E-\Omega_H^{ex} J- \Phi_q^{ex} Q_q- \Phi_p^{ex} Q_p - \Theta_H^{ex} Q_\alpha) r_0/\lambda,\nonumber\
\end{equation}
\begin{equation}
n_L = J. \label{eq:nRnLnew}
\end{equation}
Now the density matrix possesses the Boltzmann weighting factor
\begin{equation}
e^{-\left( \frac{E - \Omega_H^{ex} J - \Phi_q^{ex} Q_q- \Phi_p^{ex} Q_p - \Theta_H^{ex} Q_\alpha}{T_H}\right) } =e^{-\frac{n_R}{T_R}-\frac{n_L}{T_L}-\frac{Q_q}{T_q}-\frac{Q_p}{T_p} -\frac{Q_\alpha}{T_\alpha} }. \label{eq:Boltzmannweightingfactor}\
\end{equation}
All temperatures of the CFT are then computed as
\begin{equation}
T_R = \frac{T_H r_0}{\lambda}\bigg|_{ex} , ~~~ T_L = - \frac{\partial T_H/\partial \hat{r}_+}{\partial \Omega_H / \partial \hat{r}_+}\bigg|_{ex}, \
\end{equation}
\begin{equation}
T_q = - \frac{\partial T_H/\partial \hat{r}_+}{\partial \Phi_q / \partial \hat{r}_+}\bigg|_{ex}, ~~~ T_p = - \frac{\partial T_H/\partial \hat{r}_+}{\partial \Phi_p / \partial \hat{r}_+}\bigg|_{ex}, ~~~ T_\alpha = - \frac{\partial T_H/\partial \hat{r}_+}{\partial \Theta_H / \partial \hat{r}_+}\bigg|_{ex}.\ \label{eq:generalCFTtemperature}
\end{equation}
The right-moving temperature vanishes in the extremal limit and others remain non-zero. All non-zero CFT temperatures are given as
\begin{equation}
T_L = \frac{Vr_0^2}{4\pi a r_+}, ~~ T_\alpha = \frac{2V}{\pi r_0^2(\upsilon -1)r_+^{\upsilon -2}}, \label{eq:CFTtemperatures}
\end{equation}
\begin{equation}
T_q = \frac{Vr_0^2}{2\pi e (r_+^2 - a ^2)}, ~~ T_p =\frac{V r_0^2}{2\pi g (r_+^2 - a ^2)}. \label{eq:CFTtemperatures1}
\end{equation}
We can write the left-moving temperature as $ T_L = 1/(2\pi e )$ where $ e $ is a constant that is given in (\ref{eq:p}).  It is worth pointing out that all CFT temperatures are proportional with constant $ V $. From those temperatures, it is obvious that the dual of this black hole is described by the CFT in the mixed state. On the other hand, the quantum fields outside the horizon are not in a pure state anymore portrayed by the density matrix
\begin{equation}
\hat{\rho} = e^{-\frac{n_L}{T_L}-\frac{Q_q}{T_q}-\frac{Q_p}{T_p} -\frac{Q_\alpha}{T_\alpha}}.
\end{equation}

\subsection{Cardy Entropy}
The upshot of our calculation is to obtain the entropy formula for the near-horizon extremal rotating charged black hole surrounded by quintessence in Rastall gravity. In order to find it, we employ the famous Cardy formula
\begin{equation}
S_{CFT} = \frac{\pi^2}{3}(c_L T_L + c_R T_R). \label{eq:Cardyfromula}
\end{equation}
This entropy formula implies that the entropy of a unitary CFT is satisfied at large $ T_{R,L} $ or $ T_{R,L}\gg c_{R,L} $. Upon the insertion of the central charge (\ref{eq:cL}) and left-moving temperature in (\ref{eq:CFTtemperatures}), therefore we obtain the following entropy
\begin{equation}
S_{CFT} = \pi (r_+^2 + a^2). \label{eq:entropyquintessence}\
\end{equation}

It is important to note that from our calculation using the Kerr/CFT correspondence, we find that there is no correction regarding the existence of Rastall coupling constant as given on Eq. (\ref{eq:Scorrected}). Our result precisely matches with the Bekenstein-Hawking entropy (\ref{en1}) and supports the result given in \cite{ChakrabortyRastallEnt2018}. This also proves that the rotating charged black hole surrounded by quintessence is holographically dual with CFT. This entropy is suitable for general value of $ \kappa\lambda $ and $ \omega $. The contribution of those constants is stored in the event horizon $ r_+ $. In the next section, we are going to give several examples for specific values of $ \kappa\lambda $ and $ \omega $.

\section{Specific Black Holes Surrounded by Quintessence}

In this part, the central charge, temperature and entropy of several black hole solutions for specific values of  $ \kappa\lambda $ and $ \omega $ will be given. Recall that  $ \kappa\lambda $ is a constant measuring the deviation from Einstein gravity while $ \omega $ is to determine the matter domination around the black holes. Those constants also determine the number of the horizons. It is important since we are investigating the extremal black hole solutions with only one horizon. The number of the horizon can be identified from the constant $ \upsilon $. We narrow our calculation for $ 0 \leq \upsilon \leq 4 $ since it can be argued that the cosmological and quintessential horizons may exist beside the inner and outer horizons. Several circumstances for $ \kappa\lambda $ and $ \omega $ are shown in Table \ref{tab1}. Nevertheless, in fact, we may extend mathematically for $ \upsilon > 4 $.  We also exhibit several conditions for $ \upsilon = 1/2 $ and $ \upsilon = 3/2 $ because both values will still yield quartic roots of $ \Delta $. This still corresponds to the existence of cosmological and quintessential horizons. Below, we delve the entropy for those peculiar circumstances.

\begin{table}[!b]
    \caption{Value of $ \upsilon $ for several specific $ \kappa\lambda $ and $ \omega $.}
    \begin{minipage}{.33\linewidth}
      \centering
        \begin{tabular}{|c|c|c|}
    \hline $\upsilon$  & $\kappa\lambda$ & $ \omega $  \\ 
	\hline
	\hline 0 & any & 1/3 \\
	\hline 1 & -1/2 & -1/3 \\
	 & -2 & -2/3 \\
	 & 4 & -4/3 \\
	\hline 2 & 1/2 & 1/3 \\
	 & 1/6 & 0 \\
	 & 0 & -1/3 \\
	 & -1/2 & -2/3 \\
	 & 3/2 & -4/3 \\
	 \hline
        \end{tabular}
    \end{minipage}%
    \begin{minipage}{.33\linewidth}
      \centering
        \begin{tabular}{|c| c|c|}
    \hline $\upsilon$  & $\kappa\lambda$ & $ \omega $  \\ 
	\hline
		\hline 3 & 1/2 & 1/3 \\
	 & 2/9 & 0 \\
	 & 1/6 & -1/3 \\
	 & 0 & -2/3 \\
	 & 2/3 & -4/3 \\
	 \hline 4 & 1/2 & 1/3 \\
	 & any & -1 \\
	 & -1/4 & -4/3  \\
	\hline
        \end{tabular}
    \end{minipage} 
    \begin{minipage}{.33\linewidth}
      \centering
        \begin{tabular}{|c|c|c|}
    \hline $\upsilon$  & $\kappa\lambda$ & $ \omega $  \\ 
	\hline
	\hline 1/2 & 3/4 & 1/3 \\
	 & -1 & 0 \\
	 & -9/2 & -1/3 \\
	 & -17 & -2/3 \\
	 & 27 & -4/3 \\
	 \hline 3/2 & 3/4 & 1/3 \\
	 & 1/3 & 0 \\
	 & -1/2 & -1/3 \\
	 & -3 & -2/3 \\
	 & 7 & -4/3 \\
	 \hline
        \end{tabular}
    \end{minipage}%
    \label{tab1}   
\end{table}


\subsection{Extremal Black Hole for $ \upsilon =0, 1, 2 $}

As shown on the left part of Table \ref{tab1}, we can obtain $ \upsilon =0, 1, 2 $ for several conditions of $ \kappa\lambda  $ and $ \omega =1/3 $. This has been studied also before in \cite{SaktiEPJPlus2019}, yet for the quintessential black holes in Einstein gravity $ (\kappa\lambda =0) $ with non-vanishing NUT charge, although the resulting central charges, temperatures and entropies are same. Moreover, we can also obtain it from Eqs. (\ref{eq:cL}), (\ref{eq:CFTtemperatures}) and (\ref{eq:entropyquintessence}). The event horizons for those circumstances are located at
\begin{equation}
r_+^{(0)} = M, ~~~ r_+^{(1)} = M + \frac{\alpha}{2}, ~~~ r_+^{(2)} = \frac{M}{1-\alpha}.
\end{equation}
The upshots for the conditions in the left part of Table \ref{tab1} are given as follows
\begin{equation}
c_L^{(0)} = 12aM, ~~~c_L^{(1)} = 12a\left(M+\frac{\alpha}{2}\right), ~~~c_L^{(2)} = \frac{12aM}{(1-\alpha)^2}, \label{eq:cL012}\
\end{equation}
for the central charges, 
\begin{equation}
T_L^{(0)} = \frac{M^2 +a^2}{4\pi a M}, ~~~T_L^{(1)} = \frac{\left(2M + \alpha \right)^2 +4a^2}{8\pi a\left(2M + \alpha \right)}, \nonumber\
\end{equation}
\begin{equation}
T_L^{(2)} = \frac{M^2 +a^2(1-\alpha)^2}{4\pi a M}, \label{eq:TL012}\
\end{equation}
for the CFT temperatures, and entropies
\begin{equation}
S^{(0)}_{CFT} = \pi \big(M^2 + a^2 \big) ,~~~ S^{(1)}_{CFT} = \frac{\pi}{4} \bigg[\left(2M +\alpha\right)^2 + 4a^2 \bigg],\nonumber\
\end{equation}
\begin{equation}
S^{(2)}_{CFT} = \frac{\pi\big[ M^2 + a^2(1-\alpha)^2 \big]}{(1-\alpha)^2} .\label{eq:SCFT012}\
\end{equation}
Note that for $ S^{(2)}_{CFT} $, $ \alpha $ should be less than one to produce finite entropy.

\subsection{Extremal Black Hole for $ \upsilon =3,4 $ and $ \upsilon =1/2, 3/2 $}

Not as easy as the cases of $ \upsilon =0,1,2 $, it is quite difficult to determine the location of the single horizon (event horizon) of the extremal black holes with the initial three or four horizons, for instance when $ \upsilon =3,4 $ and $ \upsilon =1/2, 3/2 $. It is because the roots of $ \Delta $ for those values are also mathematically complicated. So, in order to avoid those complex mathematical forms, we just denote the event horizon as $ r_+ $. Using Eqs. (\ref{eq:cL}) and (\ref{eq:CFTtemperatures}), it is obtained that
\begin{equation}
c^{(3)}_L = \frac{12ar_+}{1-3\alpha r_+}, ~~~ T^{(3)}_L = \frac{(1-3\alpha r_+)(r_+^2 + a^2)}{8\pi a r_+} ,
\end{equation}
for $ \upsilon =3 $ and for $ \upsilon =4 $,
\begin{equation}
c^{(4)}_L = \frac{12ar_+}{1-6\alpha r_+^2}, ~~~ T^{(4)}_L = \frac{(1-6\alpha r_+^2)(r_+^2 + a^2)}{8\pi a r_+}.
\end{equation}
Note that $ \upsilon =1/2, 3/2 $ also yields four horizons. In these circumstances, we find
\begin{equation}
c^{(1/2)}_L = \frac{96ar_+}{8+\alpha r_+^{-3/2}}, ~~~ T^{(1/2)}_L = \frac{(8+\alpha r_+^{-3/2})(r_+^2 + a^2)}{32\pi a r_+} ,
\end{equation}
for $ \upsilon =1/2 $ and
\begin{equation}
c^{(3/2)}_L = \frac{96ar_+}{8-3\alpha r_+^{-1/2}}, ~~~ T^{(3/2)}_L = \frac{(8-3\alpha r_+^{-1/2})(r_+^2 + a^2)}{32\pi a r_+},
\end{equation}
for $ \upsilon =3/2 $. All entropies are still in the form given on Eq. (\ref{eq:entropyquintessence}).

\subsection{Extremal Quintessential Black Hole in Einstein Gravity}

Our calculation within this paper also generalizes the results in \cite{SaktiEPJPlus2019} where therein it has been taken that $ \kappa \lambda =0 $ (Einstein gravity) for $ \omega =1/3, 0, -1/3 $. The general results for any $ \omega $ in Einstein gravity can be gained also from Eqs. (\ref{eq:cL}), (\ref{eq:CFTtemperatures}) and (\ref{eq:entropyquintessence}). Using those results, we may show that for general $ \omega $ of rotating quintessential black hole
\begin{equation}
c^\omega_L = \frac{24ar_+}{2+3\alpha\omega(1-3\omega)r_+^{-1-3\omega}}, 
\end{equation}
\begin{equation}
 T^\omega_L = \frac{\left(2+3\alpha\omega(1-3\omega)r_+^{-1-3\omega}\right)(r_+^2 +a^2)}{8\pi a r_+}.
\end{equation}
Note that the entropy is given as on Eq. (\ref{eq:entropyquintessence}).

\section{Twofold CFT Duals on Extremal Black Hole}

When the Kerr black hole is coupled with the electromagnetic field, there is a possibility to extend the computation to the alternative dual CFT. It is inspired by the condition when $ a \rightarrow 0 $ that will produce a singular left-moving left temperature and zero central charge. Since it is not regular, we require a way to regularize it. In fact, we can uplift the solution when $ a \rightarrow 0 $ to a higher dimensional solution as given in Kaluza-Klein theory. In this fashion, the electromagnetic potential behaves as a new compact extra dimension with an additional fibered coordinate \cite{Hartman2009}. This alternative CFT can also be named as the second dual CFT. It is worth noting that in \cite{Ghodsi2010}, the authors have shown that the uplifted 5D metric satisfies 5D Einstein-Maxwell system where in 4D, it represents Einstein-Maxwell-dilaton theory.

Hartman \textit{et al.} \cite{Hartman2009} have managed to acquire the finite microscopic entropy by employing the second dual CFT. It has been followed in several papers \cite{Sakti2018,SaktiEPJPlus2019} as we will adopt here. Firstly, it is required to enhance the metric with a new coordinate representing the fifth dimension of $ S ^1 $ which possesses periodic property as $ z \sim z + 2 \pi R_n $. It adds a new Killing vector $ \partial_z $ which corresponds to $ U (1) $ symmetry in addition to $ SL (2, R)_R \times U (1)_L $. Moreover, as we mentioned before, the electromagnetic potential is mapped as a compact extra dimension. However, the electromagnetic potential is not regular too when it takes $ a\rightarrow 0 $. It is not a big deal since we may use a certain gauge transformation to remove the singular part. 

It is worth to point out that we can have twofold CFT dual in the alternative dual CFT due to the existence of the electromagnetic potential (\ref{eq:electromagneticpotentialKNQR}) and its dual (\ref{eq:dualelectromagneticpotentialKNQR}). Hence, even for vanishing quintessential matter, we could still investigate two different pictures that we call as $ Q $-picture and $ P $-picture within this paper. Those pictures are linked to the mapping of the electromagnetic potential and its dual to enhance 5D metric.

\subsection{$ Q $-Picture}

The first picture is $ Q $-picture where it refers to the mapping of (\ref{eq:nearhorizonelectromagneticpot1}) to enhance the 5D metric. In the limit $ a\rightarrow 0 $, we may apply the following gauge transformation
\begin{eqnarray}
\textbf{A} \rightarrow \textbf{A} - \frac{qr_+}{2a} d\varphi ,\
\end{eqnarray}
to remove the singularity on the electromagnetic potential. Following above transformation, the electromagnetic potential transforms as
\begin{eqnarray}
\textbf{A}=\frac{qr}{V}  d\tau + p\cos\theta d\varphi . \label{eq:elecpotRN} \
\end{eqnarray}
Now the uplifted metric is given in the form
\begin{eqnarray}
ds_5^2 = ds_4^2 + (dz + \textbf{A})^2, \label{eq:RN5D} \
\end{eqnarray}
which is a 5D Reissner-Nordstr{\"o}m black hole solution surrounded by quintessence. In this uplifted space-time metric, the electromagnetic potential contains an angular velocity which corresponds to the electric charge $ q $. The four-dimensional metric is given by
\begin{equation}
ds_4^2= \frac{r_+^2}{V} \left(- r^2 d\tau ^2 + \frac{dr^2}{r^2}+V d\theta ^2 + V \sin^2\theta d\varphi^2 \right). \label{eq:RNQ4D}
\end{equation}
It is worth noting that this computation might not be valid in Rastall gravity since we do not know the Kaluza-Klein dimensional reduction for Rastall gravity. So, we switch off $ \kappa\lambda $ to meet the Kaluza-Klein theory's requirement. The second dual CFT is also aimed to prove that the microscopic entropy matches with the Bekenstein-Hawking one. For this solution, the Bekenstein-Hawking entropy is given as
\begin{eqnarray}
S_{BH} = \pi r_+^2 = \frac{A_H}{4G_5}.\label{eq:entropy5D}
\end{eqnarray}

The central charge is computed again using ASG. To look for the non-trivial diffeomorphisms, we need to set some consistent boundary conditions of the metric deviation as in 4D solution. Herein the same boundary conditions such in  \cite{Hartman2009,Sakti2018,SaktiEPJPlus2019} are adopted, i.e.
\begin{eqnarray}
h_{\mu \nu} \sim \left(\begin{array}{ccccc}
\mathcal{O}(r^2) & \mathcal{O}\left(\frac{1}{r^2}\right) &  \mathcal{O}\left(\frac{1}{r}\right) & \mathcal{O}(r) &  \mathcal{O}(1) \\
 &  \mathcal{O}\left(\frac{1}{r^3}\right) &  \mathcal{O}\left(\frac{1}{r^2}\right) & \mathcal{O}\left(\frac{1}{r}\right) &  \mathcal{O}\left(\frac{1}{r}\right) \\
 &  & \mathcal{O}\left(\frac{1}{r}\right) & \mathcal{O}(1) &  \mathcal{O}\left(\frac{1}{r}\right)\\
 &  &  & \mathcal{O}\left(\frac{1}{r}\right) & \mathcal{O}(1)\\
 &  &  &  & \mathcal{O}(1)\
\end{array} \right), \label{eq:gdeviation2}
\end{eqnarray}
in the basis $ (\tau,r,\theta, \varphi, z) $. Those boundary conditions (\ref{eq:gdeviation2}) allow
\begin{eqnarray}
\zeta_z = \epsilon(z) \partial_z - r \epsilon '(z)\partial _z ,\ \label{eq:difeoz}
\end{eqnarray}
but do not allow $ \zeta_\epsilon $ such the case of 4D Kerr-Newman-NUT black holes in which $ \epsilon (z) = -e^{-inz} $ and the prime $ (') $ denotes the derivative with respect to $ z $. The central charge can be obtained from the 5D generalization of the treatment in \cite{Hartman2009}. It turns out that the central charge is given by
\begin{eqnarray}
c^{(q)}_z = \frac{6qr^2_+}{V}, 
\end{eqnarray}
that is associated to $ \zeta_z $.

To further calculate the entropy, we require the CFT temperatures. Since $ T_L \rightarrow 0 $, there are left only the temperatures conjugate to the electric, magnetic and quintessential charges. By employing the first law of black hole thermodynamics, we now possess
\begin{eqnarray}
dS = \frac{dQ_q}{T_q}+\frac{dQ_p}{T_p} +\frac{Q_\alpha}{T_\alpha}.
\end{eqnarray}
We are dealing with $ Q $-picture, so we demand the temperature conjugate to the electric charge. Taking $ a\rightarrow 0 $ in the temperature (\ref{eq:CFTtemperatures1}), then we find
\begin{equation}
T_q = \frac{Vr_0^2}{2\pi q r_+^2}.
\end{equation}

In the end of our computation, we are able to compute the microscopic entropy of the extremal Reissner-Nordstr{\"o}m black hole surrounded by quintessence in $ Q $-picture by governing Cardy formula. The entropy of this black hole is then given as follows
\begin{eqnarray}
S_{CFT} = \frac{\pi ^2}{3}c^{(q)}_z T_q = \pi r_+^2 = \frac{A_H}{4G_5}.
\end{eqnarray}
Therefore the entropy from the CFT for this 5D solution also matches with the Bekenstein-Hawking entropy.

\subsection{$ P $-Picture}

The second picture is $ P $-picture where it refers to the mapping of (\ref{eq:nearhorizonelectromagneticpot2}) to enhance the 5D metric. So, the dual electromagnetic potential will be the source of rotation in the uplifted space-time metric. The following gauge transformation can be applied
\begin{eqnarray}
\bar{\textbf{A}} \rightarrow \bar{\textbf{A}} - \frac{pr_+}{2a} d\varphi .\
\end{eqnarray}
as well as taking the limit $ a\rightarrow 0 $ to remove the singularity term on the electromagnetic potential. After the transformation, the electromagnetic potential is given as follows
\begin{eqnarray}
\bar{\textbf{A}} =\frac{pr}{V}  d\tau -q\cos\theta d\varphi . \label{eq:elecpotRNNUT} \
\end{eqnarray}
Now the uplifted metric is given in the form
\begin{eqnarray}
ds_5^2 = ds_4^2 + (dz + \bar{\textbf{A}})^2. \label{eq:RN5dual} \
\end{eqnarray}
The four-dimensional metric is also given by Eq. (\ref{eq:RNQ4D}). The Bekenstein-Hawking entropy of this 5D solution is also given on Eq. (\ref{eq:entropy5D}).

By employing the similar boundary conditions as given on Eq. (\ref{eq:gdeviation2}), one can obtain an identical diffeomorphism as Eq. (\ref{eq:difeoz}). Finally, the central charge related to metric (\ref{eq:RN5dual}) and associated to $ \zeta_z $ is given by
\begin{eqnarray}
c^{(p)}_z =\frac{6pr_+^2}{V} .
\end{eqnarray}

In this $ P $-picture, we implement the temperature conjugate to the magnetic charge. This is indicated by Eq. (\ref{eq:CFTtemperatures1}). Hence, for $ a\rightarrow 0 $, we have
\begin{equation}
T_p = \frac{Vr_0^2}{2\pi p r_+^2}.
\end{equation}

Finally, we are able to compute the microscopic entropy of the solution (\ref{eq:RN5dual}) using Cardy formula. The CFT entropy is then given by
\begin{eqnarray}
S_{CFT} = \frac{\pi ^2}{3}c^{(p)}_z T_p =\pi r_+^2 = \frac{A_H}{4G_5}.
\end{eqnarray}
It is found that the CFT entropy of $ P $-picture matches exactly with the entropy in $ Q $-picture and Bekenstein-Hawking entropy. Hence, we have demonstrated that there exists twofold dual CFT represented by $ Q $- and $ P $-pictures on 5D Reissner-Nordstr{\"o}m black hole surrounded by quintessence.

\section{Conclusions}\label{conc}

We have explicitly performed the Kerr/CFT duality to the black hole solution in Rastall gravity. To be specific, we have extended the Kerr/CFT duality to compute the entropy for the rotating charged black holes surrounded by quintessence in Rastall gravity. The contribution to the central charge only comes from the space-time metric since the gravitational Lagrangian is coincident with the Einstein general relativity. From our calculation, it has been shown that there is no entropy correction related to the non-vanishing Rastall coupling constant $ \kappa\lambda $. We have also given the upshots of the central charges and temperatures for several specific values of $ \kappa\lambda $ and $ \omega $ that associate with the number of the horizons of the black hole. Indeed, all entropies from CFT match with the Bekenstein-Hawking formula. Hence, this black hole solution is holographically dual to 2D CFT.

To continue supporting the holograph, we have extended the duality to the alternative dual CFT when the spin vanishes.
The alternative dual CFT has been performed on the solution obtained using the Kaluza-Klein dimensional reduction. The resulting space-time is 5D Reissner-Nordstr{\"o}m black hole surrounded by quintessence. We have switched off $ \kappa\lambda $ for this computation because the Kaluza-Klein dimensional reduction for Rastall gravity is still unclear. We have benefited that since the electromagnetic potential possesses its dual, we could construct 5D Reissner-Nordstr{\"o}m black hole surrounded by quintessence in $ Q $- and $ P $- pictures related to the electromagnetic charges that play role as the angular velocity of the black hole. By using the Kerr/CFT duality, both pictures exactly produce the identical microscopic entropies that match with the Bekenstein-Hawking entropy.

A very interesting work for the future is to examine the extended family of the hidden conformal symmetry for the rotating charged black hole solution surrounded by quintessence in Rastall gravity. We may also deform the wave equation to further calculate the effect on the conformal symmetry.

\section*{Acknowledgments}
This work is supported by Hibah Penelitian World Class University Postdoctoral Program, Institut Teknologi Bandung 2020.


\appendix
\section{Rastall Gravitational Theory from $ F(R,T) $ Gravity}\label{app:Rastalltheor}

In Einstein theory and most of the other gravitational theories, the source of the matter is portrayed by a minimal coupling between geometry and matter. Hence, the energy-momentum tensor complies with the conservation law. However, the coupling between geometry and matter can occur to be non-minimal as a general condition. It will violate the conservation law of the energy-momentum tensor. Rastall gravity is one example that allows the non-conserved covariant derivative of energy-momentum tensor. In this theory, the covariant derivative of energy-momentum tensor satisfies the following condition \cite{Rastall1972,Rastall1976}
\begin{equation}
\nabla_\mu T^{\mu\nu}=\lambda \nabla^{\nu} R, \label{eq:nonconserved}
\end{equation}
where $ \lambda $ is a coupling constant to quantify the deviation to Einstein theory of gravity. $ R $ is Ricci scalar. This is the basic condition of Rastall theory although one can generalize by replacing $ R \rightarrow F(R)$ where the function $ F(R) $ can be a non-linear function of Ricci scalar. For flat space-time, Eq. (\ref{eq:nonconserved}) will be obviously satisfied. Employing (\ref{eq:nonconserved}) to the ordinary Einstein field equation, one can obtain
\begin{equation}
G_{\mu\nu} + \kappa \lambda g_{\mu\nu} R = \kappa T_{\mu\nu}. \label{eq:Rastalleq}
\end{equation}
where $ G_{\mu\nu}= R_{\mu\nu} - \frac{1}{2}g_{\mu\nu}R $ and $ \kappa $ is the Rastall gravitational constant. Using a little bit of algebra, one can find $ R= \frac{\kappa  }{4\kappa \lambda -1} T$. For vanishing $\lambda$, it will reduce to the ordinary Einstein field equation. From Eq. (\ref{eq:nonconserved}) and $ R= \frac{\kappa T}{4\kappa \lambda -1} $, one can obtain
\begin{equation}
\nabla_\mu T^{\mu\nu}=\frac{\kappa \lambda}{4\kappa\lambda -1} \nabla^{\nu} T. \label{eq:nonconserved1}
\end{equation}

In the first formulation of Rastall gravity, Rastall \cite{Rastall1972,Rastall1976} has formulated the equation of motion without describing the action. Nonetheless, in \cite{Fisher2019,Moraes2019}, Rastall gravity is believed to be a specific case of $ F(R,T) $ gravitational theory or $ F(R,\mathcal{L}_m) $ where $ \mathcal{L}_m $ is a matter Lagrangian. Moreover, authors in \cite{ShabaniZiaieEPL2020} take specific matter Lagrangian to relate Rastall gravity and $ F(R,T) $ theory. 

It is believed that Rastall gravity can be constructed from the following action \cite{ShabaniZiaieEPL2020}
\begin{equation}
\int d^4 x \left(F(R,T) + \mathcal{L}_m \right) =\int d^4 x \left( R+\beta T + \mathcal{L}_m \right),
\end{equation} 
where $ \beta$ is a coupling constant that will be computed later and matter Lagrangian is chosen as $ \mathcal{L}_m = \bar{p} $ as pressure for a perfect fluid. Varying above action with respect to the metric, one can find the following field equation
\begin{equation}
G_{\mu\nu} = \left(\kappa' +\beta \right)T_{\mu\nu} + \beta\left(\frac{1}{2}T - \bar{p} \right), \label{eq:FRT1}
\end{equation}
where $ \kappa ' $ is the usual Einstein gravitational constant. Considering the perfect fluid equation of state $ \bar{p} = \omega_p \rho $ will lead to $ T =-\rho +3\bar{p} = (3\omega_p - 1)\rho $, so we obtain
\begin{equation}
G_{\mu\nu} = \left(\kappa' +\beta \right)T_{\mu\nu} + \frac{\beta(\omega_p -1)}{2(3\omega_p -1)}Tg_{\mu\nu}. \label{eq:FRT2}
\end{equation}
By comparing Eq. (\ref{eq:Rastalleq}) and Eq. (\ref{eq:FRT2}), one can get
\begin{equation}
\kappa  = \kappa ' + \beta, ~~~\text{where}~~~ \beta = \frac{2(1-3\omega_p)\lambda \kappa^2}{(\omega_p -1)(4\kappa  \lambda -1)} .\label{eq:relationalpha}
\end{equation}
This yields the following non-conservative matter tensor,
\begin{equation}
\nabla_\mu T^{\mu\nu} = \frac{\beta (\omega_p -1)}{2(\kappa ' +\beta)(1-3\omega_p)} \nabla^{\nu} T =\frac{\kappa \lambda}{4\kappa \lambda -1} \nabla^{\nu} T. \label{eq:nonconserved2}
\end{equation}
Therefore, in order to construct Rastall gravitational theory from $ F(R,T) $ gravity, we demand to distinguish the Rastall gravitational constant $ \kappa $ and Einstein gravitational constant $ \kappa ' $.

\section{Kerr-Newman-AdS Black Hole Surrounded by Quintessence in Rastall Gravity}\label{app:KNUTAdSQ}
The authors in \cite{XuBHRastallAdS2018} claim that they have derived the Kerr-Newman-AdS solution surrounded by a perfect fluid matter (quintessence) in Rastall gravity. The space-time metric they have obtained is given by
\begin{eqnarray}
ds^2 &=& - \frac{\Delta}{\varrho^2 \Xi} \bigg( dt - a \sin^2\theta d\phi \bigg)^2 +  \frac{\varrho ^2}{\Delta}d\hat{r}^2 + \frac{\varrho ^2}{\Delta_\theta} d\theta^2 \nonumber\\
&& + \frac{\Delta_\theta \sin^2\theta}{\varrho ^2 \Xi}\bigg[a dt - (\hat{r}^2+a^2 ) d\phi \bigg]^2,\label{KNQAdSRmetric} \
\end{eqnarray}
where
\begin{equation}
\Delta =\hat{r}^2- 2M \hat{r} +a^2 + q^2 -\alpha \hat{r}^\upsilon -\frac{\Lambda}{3}\hat{r}^2(\hat{r}^2+a^2), \nonumber\ \label{DelAdS}
\end{equation}
\begin{equation}
\Delta_\theta = 1+ \frac{\Lambda}{3}a^2\cos^2\theta ,\label{DelAdS} \nonumber\
\end{equation}
\begin{equation}
\upsilon = \frac{1-3\omega}{1-3\kappa\lambda(1+\omega)}, ~~~ \varrho^2 =\hat{r}^2+ a^2\cos^2\theta, ~~~\Xi =1+\frac{\Lambda}{3}a^2 .\label{rhoAdS} \nonumber\
\end{equation}
$ \Lambda $ stands for the cosmological constant. It is easy to derive a rotating axisymmetric solution using Newman-Janis algorithm. However, to add the cosmological constant as well as the spin, that algorithm cannot be implemented. In their derivation, the authors actually have guessed the metric solution and calculate the non-vanishing components of the Einstein tensor to prove that the metric is the solution to Rastall field equation for non-vanishing $ \Lambda $ and $ a $.  They have also assumed $ p=0 $. When $ \Lambda = 0 $, their solution matches with the solution obtained using Newman-Janis algorithm. Yet for $ a = 0 $, we find that their solution does not match with the following metric
\begin{equation}
ds^2 = -f(\hat{r}) dt^2 +\frac{d\hat{r}^2}{f(\hat{r})} + \hat{r}^2\left(d\theta^2 +\sin^2\theta d\phi^2 \right),
\end{equation}
where
\begin{equation}
f(r) = 1-\frac{2M}{\hat{r}}+\frac{q^2}{\hat{r}^2} -\frac{\Lambda}{3-12\kappa\lambda}\hat{r}^2 -\alpha \hat{r}^{-\frac{1+3\omega-6\kappa\lambda(1+\omega)}{1-3\kappa\lambda(1+\omega)}}
\end{equation}
of which this is the solution to the Rastall field equation with a cosmological constant term, electromagnetic field and quintessential field. Hence, the authors \cite{XuBHRastallAdS2018} should clarify about their solution. In addition, in \cite{WangSciChina2019} there is a derivation of the Kerr-Newman-AdS black hole solution surrounded by quintessence using the constraint from the first law of thermodynamics \footnote{Nonetheless, the authors in \cite{WangSciChina2019} do not promote the quintessential intensity as a thermodynamic quantity.} and they have performed that the solution in \cite{Wang2017a} (when $ \kappa\lambda =0 $) does not satisfy the weak energy condition. So, the solution given on Eq. (\ref{KNQAdSRmetric}) will also violate the weak energy condition. 



 \bibliographystyle{elsarticle-harv}

\begin{thebibliography}{}
%
%

\bibitem{Copeland2006}
E. J. Copeland, M. Sami and S. Tsujikawa, Int. J. Mod. Phys. D \textbf{15} (2006) 1753 .

\bibitem{Kiselev2003}
V. V. Kiselev, Class. Quantum Grav. \textbf{20} (2003) 1187.

\bibitem{Maharaj2017}
S. G. Ghosh, M. Amir and S. D. Maharaj, Eur. Phys. J. C  \textbf{77} (2017) 530.

\bibitem{Lee2018}
S. G. Ghosh, S. D. Maharaj, D. Baboolal and T. Lee, Eur. Phys. J. C \textbf{78} (2018) 90.

\bibitem{SaktiarXivString2019}
M. F. A. R. Sakti, H. L. Prihadi, A. Suroso and F. P. Zen, arXiv:1911.07569.

\bibitem{Rastall1972}
P. Rastall, Phys. Rev. D \textbf{6} (1971) 3357.

\bibitem{Rastall1976}
P. Rastall, Can. J. Phys. \textbf{54} (1976) 66.

\bibitem{Fisher2019}
S. B. Fisher and E. D. Carlson, Phys. Rev. D \textbf{100} (2019) 064059. 

\bibitem{Moraes2019}
W. A. G. De Moraes and A. F. Santos, Gen. Relativ. Grav. \textbf{51} (2019) 167. 

\bibitem{ShabaniZiaieEPL2020}
H. Shabani and A. H. Ziaie, EPL \textbf{129} (2020)  20004.

\bibitem{Campos2013}
J. P. Campos, J. C. Fabris, R. Perez, O. F. Piattella and H. E. S. Velten, Eur. Phys. J. C \textbf{73} (2013) 2357. 

\bibitem{Fabris2012}
J. C. Fabris, M. H. Daouda and O. F. Piattella, Phys. Lett. B \textbf{711} (2012) 232. 

\bibitem{Moradpour2016}
H. Moradpour, Phys. Lett. B \textbf{757} (2016) 187-191. 


\bibitem{Rawaf1996}
A. S. Al-Rawaf and M. O. Taha, Gen. Relativ. Grav. \textbf{28} (1996) 935.

\bibitem{Batista2012}
C. E. M. Batista, M. H. Daouda, J. C. Fabris, O. F. Piattella and D. C. Rodrigues, Phys. Rev. D \textbf{85} (2012) 084008.

\bibitem{Carames2014}
T. R. P. Caram{\^e}s, M. H. Daouda, J. C. Fabris, A. M. Oliveira, O. F. Piattella and V. Strokov, Eur. Phys. J. C \textbf{74} (2014) 3145. 

\bibitem{Salako2016}
I. G. Salako, M. J. S. Houndjo and A. Jawad, Int. J. Mod. Phys. D \textbf{25} (2016) 1650076. 

\bibitem{Smalley1974}
L. L. Smalley, Phys. Rev. D \textbf{9} (1974) 1635.

\bibitem{Smalley1975}
L. L. Smalley, Phys. Rev. D \textbf{12} (1975) 376.

\bibitem{Wolf1986}
C. Wolf, Phys. Scr. \textbf{34} (1986) 193.

\bibitem{Moradpour2017}
H. Moradpour, Y. Heydarzade, F. Darabi, I. G. Salako, Eur. Phys. J. C \textbf{77} (2017) 259. 

\bibitem{Hansraj2019}
S. Hansraj, A. Banerjee and P. Channuie, Ann. Phys. \textbf{400} (2019) 320. 

\bibitem{Abbas2018} 
G. Abbas and M. R. Shahzad, Eur. Phys. J. A  \textbf{54} (2018) 211. 

\bibitem{Abbas2019}
G. Abbas and M. R. Shahzad, Astrophys. Space Sci. \textbf{364} (2019) 50. 

\bibitem{Oliveira2015}
A. M. Oliveira, H. E. S. Velten, J. C. Fabris and L. Casarini, Phys. Rev. D \textbf{92} (2015) 044020. 

\bibitem{Moradpour2017a}
H. Moradpour and N. Sadeghnezhad, Can. J. Phys. \textbf{95} (2017) 1257. 

\bibitem{Moradpour2016a}
H. Moradpour and I. G. Salako, Adv. High Energy Phys. \textbf{2016} (2016) 3492796. 

\bibitem{Oliveira2016}
A. M. Oliveira, H. E. S. Velten and J. C. Fabris, Phys. Rev. D \textbf{93} (2016) 124020. 

\bibitem{Bronnikov2016}
K. A. Bronnikov, J. C. Fabris, O. F. Piattella and E. C. Santos, Gen. Relativ. Grav. \textbf{48} (2016) 162.

\bibitem{SaktiAnnPhys2020}
M. F. A. R. Sakti, A. Suroso and F. P. Zen, Ann. Phys. \textbf{413} (2020) 168062. 

\bibitem{HadyanIJMPD2020}
H. L. Prihadi, M. F. A. R. Sakti, G. Hikmawan and F. P. Zen, Int. J. Mod. Phys. D \textbf{29} (2020) 2050021. 

\bibitem{Guica2009}
M. Guica, T. Hartman, W. Song and A. Strominger, Phys. Rev. D \textbf{80} (2009) 124008.

\bibitem{Hartman2009}
T. Hartman, K. Murata, T. Nishioka and A. Strominger, JHEP \textbf{04} (2009) 019. 

\bibitem{Ghezelbash2009}
A. M. Ghezelbash, JHEP \textbf{08} (2009) 045.

\bibitem{Lu2009}
H. L{\"u}, J. Mei and C. N. Pope, JHEP \textbf{04} (2009) 054.

\bibitem{Li2010}
R. Li and J-R. Ren, JHEP \textbf{09} (2010) 039. 

\bibitem{Anninos2010}
D. Anninos and T. Hartman, JHEP \textbf{03} (2010) 096. 

\bibitem{Ghodsi2010}
A. Ghodsi and M. R. Garousi, Phys. Lett. B \textbf{687} (2010) 79. 

\bibitem{Ghezelbash2012}
A. M. Ghezelbash, Mod. Phys. Lett. A \textbf{27} (2012) 1250046. 

\bibitem{Astorino2015}
M. Astorino, JHEP \textbf{10} (2015) 016. 

\bibitem{Astorino2015a}
M.  Astorino, Phys. Lett. B \textbf{751} (2015) 96. 

\bibitem{Siahaan2016}
H. M. Siahaan, Class. Quantum Grav. \textbf{33} (2016) 155013. 

\bibitem{Astorino2016}
M. Astorino, Phys. Lett. B \textbf{760} (2016) 393. 

\bibitem{Sinamuli2016}
M. Sinamuli and R. B. Mann, JHEP \textbf{08} (2016) 148. 

\bibitem{Sakti2018}
M. F. A. R. Sakti, A. Suroso and F. P. Zen, Int. J. Mod. Phys. D \textbf{27} (2018) 1850109. 

\bibitem{SaktiEPJPlus2019}
M. F. A. R. Sakti, A. Suroso and F. P. Zen, Eur. Phys. J. Plus \textbf{134} (2019) 580. 

\bibitem{SaktiMicroJPCS2019}
M. F. A. R. Sakti, A. Suroso and F. P. Zen, J. Phys.: Conf. Ser. \textbf{1204} (2019) 012009. 

\bibitem{Castro2010}
A. Castro, A. Maloney and A. Strominger, Phys. Rev. D \textbf{82} (2010) 024008. 

\bibitem{ChenLong2010}
B. Chen, J. Long and J-J. Zhang, Phys. Rev. D \textbf{82} (2010) 104017. 

\bibitem{ChenSun2010}
C-M. Chen and J-R. Sun, JHEP \textbf{08} (2010) 034. 

\bibitem{ChenWang2010}
D. Chen, H. Wang, H. Wu and H. Yang, JHEP \textbf{11} (2010) 002. 

\bibitem{WangLiu2010}
Y-Q. Wang and Y-X. Liu, JHEP \textbf{08} (2010) 087. 

\bibitem{ChenLongJHEP2010}
B. Chen and J. Long, JHEP \textbf{08} (2010) 065. 

\bibitem{SetareKamali2010}
M. R. Setare and V. Kamali, JHEP \textbf{10} (2010) 074. 

\bibitem{GhezelbashKamali2010}
A. M. Ghezelbash, V. Kamali and M. R. Setare, Phys. Rev. D \textbf{82} (2010) 124051. 

\bibitem{ChenHuangPRD2010}
C-M. Chen, Y-M. Huang, J-R. Sun, M-F. Wu and S-J. Zou, Phys. Rev. D \textbf{82} (2010) 066004. 

\bibitem{ChenChen2011}
B. Chen, C-M. Chen and B. Ning, Nuc. Phys. B \textbf{853} (2011) 196-209.

\bibitem{ChenGhezelbash2011}
B. Chen, A. M. Ghezelbash, V. Kamali and M. R. Setare, Nuc. Phys. B \textbf{848} (2011) 108-120. 

\bibitem{ChenZhangJHEP2011}
B. Chen B and J-J. Zhang, JHEP \textbf{08} (2011) 114. 

\bibitem{HuangYuan2011}
Y-C. Huang and F-F. Yuan, JHEP \textbf{03} (2011) 029. 

\bibitem{Shao2011}
K-N. Shao and Z. Zhang, Phys. Rev. D \textbf{83} (2011) 106008. 

\bibitem{DeyouChen2011}
D. Chen, P. Wang, H. Wu and H. Yang, Gen. Relativ. Grav. \textbf{43} (2011) 181-190. 

\bibitem{GhezelbashSiahaan2013}
A. M. Ghezelbash and H. M. Siahaan, Class. Quantum Grav. \textbf{30} (2013) 135005. 

\bibitem{SiahaanAcc2018}
H. M. Siahaan, Class. Quantum Grav. \textbf{35} (2018) 155002. 

\bibitem{Saktideformed2019}
M. F. A. R. Sakti, A. M. Ghezelbash, A. Suroso and F. P. Zen,  Gen. Relativ. Grav. \textbf{51} (2019) 151. 

\bibitem{SaktiNucPhysB2020}
M. F. A. R. Sakti, A. M. Ghezelbash, A. Suroso and F. P. Zen,  Nuc. Phys. B \textbf{953} (2020) 114970. 

\bibitem{Compere2017}
G. Comp{\'e}re, Living Rev. Rel. \textbf{15} (2012) 11 [Addendum ibid. \textbf{20} (2017) 1]. 

\bibitem{XuBHRastallAdS2018}
Z. Xu, X. Hou, X. Gong and J. Wang, Eur. Phys. J. C \textbf{78} (2018) 513. 

\bibitem{ChenPRDQuin2008}
S. Chen, B. Wang and R. Su, Phys. Rev. D \textbf{77} (2008) 124011. 

\bibitem{Sekiwa2006}
Y. Sekiwa, Phys. Rev. D \textbf{73} (2006) 084009. 

\bibitem{ChakrabortyRastallEnt2018}
D. Das, S. Dutta and S. Chakraborty, Eur. Phys. J. C \textbf{78} (2018) 810. 

\bibitem{MoradpourRastallEnt2016}
H. Moradpour and I. G. Salako, Adv. High Energy Phys. \textbf{2016} (2016) 3492796. 

\bibitem{ChaiZhangWangCTP2005}
R. G. Cai, H. S. Zhang and A. Wang,  Commun. Theor. Phys. \textbf{44} (2005) 948.

\bibitem{ChakraPadmaPRD2014a}
S. Chakraborty and T. Padmanabhan, Phys. Rev. D \textbf{90} (2014) 084021.

\bibitem{ChakraPadmaPRD2014b}
S. Chakraborty and T. Padmanabhan, Phys. Rev. D \textbf{90} (2014) 124017.

\bibitem{MitraGRG2015}
S. Mitra, S. Saha and S. Chakraborty, Gen.
Rel. Grav. \textbf{47} (2015) 69.

\bibitem{brown1986}
J. D. Brown and M. Henneaux, Commun. Math. Phys. \textbf{104} (1986) 207. 

\bibitem{BarnichBrandt2002}
G. Barnich and F. Brandt, Nuc. Phys. B \textbf{633} (2002) 3. 

\bibitem{CompereMurataJHEP2009}
G. Comp{\'e}re, K. Murata and T. Nishioka, JHEP \textbf{05} (2009) 077. 

\bibitem{WangSciChina2019}
Y. Wang, C-H. Wu and R-H. Yue, Sci. China Phys. Mech. Astro. \textbf{62} (2019) 110411. 


\bibitem{Wang2017a}
Z. Xu and J. Wang, Phys. Rev. D \textbf{95} (2017) 064015. 

\end{thebibliography}

\end{document}